\documentclass[submission]{eptcs}
\providecommand{\event}{FMSPLE 2015} 
\usepackage{breakurl}             

\usepackage{abbrevs}
\usepackage{xspace}

\newcommand{\PL}{\ac{PL}\xspace} 
 
\newcommand{\SPLs}{\acp{SPL}\xspace}

\newcommand{\SAT}{\ac{SAT}\xspace}  
\newcommand{\CSP}{\ac{CSP}\xspace}  
  
\newcommand{\CTC}{\ac{CTC}\xspace}  
\newcommand{\CTCs}{\acp{CTC}\xspace}  
  
\newcommand{\FM}{\ac{FM}\xspace}
\newcommand{\FMs}{\acp{FM}\xspace}  
  
\newcommand{\AFM}{\ac{AFM}\xspace}  
\newcommand{\AFMs}{\acp{AFM}\xspace}  
\newcommand{\SFM}{\ac{SFM}\xspace}

\usepackage[nolist]{acronym}
\begin{acronym}

\acro{PL}{product line}
\acro{SPL}{software product line}
\acro{MDSoC}{multi-dimensional separation of concerns}
\acro{SoC}{separation of concerns}
\acro{FODA}{feature oriented domain analysis}
\acro{PDF}{portable document format}
\acro{OCR}{object character recognition}
\acro{SAT}{satisfiability}
\acro{CSP}{constraint satisfaction problem}
\acro{CTCR}{cross-tree constraint ratio}
\acro{CTC}{cross-tree constraint}
\acro{SPLOT}{software product line online tools}
\acro{FM}{feature model}
\acro{EFM}{extended feature model}
\acro{AFM}{attributed feature model}
\acro{SFM}{stateful feature model}
\acro{LOC}{lines of code}

\end{acronym}

\usepackage{balance}
\usepackage{graphics}
\usepackage{url}
\usepackage{tikz}
\usetikzlibrary{arrows,decorations.pathmorphing,backgrounds,positioning,fit,petri,shapes}
\usepackage{rotating}
\usepackage{listings}
\lstdefinestyle{numbers}
{numbers=left, stepnumber=1, numberstyle=\tiny, numbersep=10pt}
\lstdefinestyle{nonumbers}
{numbers=none}
\usepackage{float}
\usepackage{afterpage}
\usepackage{multirow}
\usepackage{caption} 
\usepackage{paralist}
\newfloat{listing}{tbp}{lol}
\floatname{listing}{Listing}
\floatstyle{boxed}
\restylefloat{listing}
\newenvironment{listingsize}{\begin{small}\normalsize\small}{\end{small}}
\lstdefinestyle{mystyle}{
    flexiblecolumns=true,
    keepspaces=true,
    basewidth={0em,0em},
    columns = fullflexible,
    fontadjust = true,
    showstringspaces = false,
    inputencoding=utf8,
    basicstyle=\scriptsize\sffamily,
    keywordstyle=\bfseries
}
\title{Detecting and Explaining Conflicts \\in Attributed Feature Models\thanks{This work was partially supported by the DFG (German Research Foundation) under grants SCHA1635/2-1 and SCHA1635/4-1 and by the European Commission within the project HyVar (grant agreement H2020-644298).}}
\author{Uwe Lesta \qquad\qquad Ina Schaefer \qquad\qquad Tim Winkelmann
\institute{Institute of Software Engineering and Automotive Informatics\\
Technische Universit{\"a}t Braunschweig\\
Braunschweig, Germany}
\email{Lesta@sbs-softwaresysteme.de \quad i.schaefer@tu-braunschweig.de \quad t.winkelmann@tu-braunschweig.de}
}
\def\titlerunning{Detecting and Explaining Conflicts in Attributed Feature Models}
\def\authorrunning{U. Lesta, I. Schaefer \& T. Winkelmann}
\date{30 Januar 2015}
\begin{document}

\maketitle
\begin{abstract}
Product configuration systems are often based on a variability model.
The development of a variability model is a time consuming and error-prone
process.
Considering the ongoing development of products, the variability model has to be
adapted frequently.
These changes often lead to mistakes, such that some products cannot be derived
from the model anymore, that undesired products are derivable or that there are contradictions in the variability model. 
In this paper, we propose an approach to discover and to explain contradictions in
attributed feature models efficiently in order to assist the developer with the
correction of mistakes. 
We use extended feature models with attributes and arithmetic constraints,
translate them into a constraint satisfaction problem and explore those for contradictions. 
When a contradiction is found, the constraints are searched for a set of
contradicting relations by the QuickXplain algorithm.
\end{abstract}

%
%
\vspace{-0.5cm}
\section{Introduction}
\label{sec:intro}
\vspace{-0.3cm}

Matys~\cite{matys2011praxishandbuch} describes a \PL as a group of related products which fulfill similar
functions, complement one another or are destined for the same target group.
Variability models are used to support the product configuration process and to analyze the variability of the existing \PL. In the context of 
\SPLs~\cite{Pohl:2005:SPL:1095605}, \FMs~\cite{Kang1990} are a widely used formalism for variability modeling. 
%
%
%
Product configuration is an optimization task in order to 
find the most suitable product that meets a set of predetermined requirements and restrictions.
The definition of an objective function for this optimization in a pure \FM, i.e., a feature model without any attributes, 
can only refer to the section of as many or as few features as possible. 
In \AFMs~\cite{DBLP:conf/caise/BenavidesTC05}, where each feature can have multiple
integer-valued attributes, optimization during product configuration becomes much more interesting.
In particular, as attributes can have arithmetic constraints between each other which need to be satisfied in a configured product.
%
Defining large and complex variability models, such as \AFMs with integer-valued
attributes and arithmetic constraints, is error-prone. 
Errors in the variability model may lead either to an undesired configurable product 
or to a realizable product which cannot be configured.
Another type of modeling error is to relate features or attribute values in a wrong
way which may either lead to a contradiction between the model and the reality
or to a contradiction within the variability model.
Contradictions within the variability model are hard to find and are mostly
accidentally discovered without tool support.
Furthermore, it is a time consuming and difficult task to find the reason for
such a contradiction manually in order to resolve it.

To counter this problem, we present an approach to detect contradictions within
\AFMs in an efficient way.
We further provide means to explain the cause of the contradiction to support the modeler to build a consistent \AFM.
There has been a lot of work on finding contradictions in the \FM community on pure \FMs~\cite{Felfernig2012,Zhang2011ICSR,Xiong2012ICSE, Trinidad12-PhD,Benavides:2010:AAF:1786808.1787031}, 
but to the best of our knowledge there were only suggestions to do this on \AFMs with integer-valued attributes and general arithmetic constraints.
The analysis of \AFMs with
integer-valued attributes and arithmetic constraints is particularly complex~\cite{Benavides:2010:AAF:1786808.1787031}, since not only attribute-value pairs have to be considered, but also arithmetic constraints.
The number of products derived from the model can change drastically if such arithmetic constraints are involved. 
We describe how an \AFM with integer attributes can
be translated into a \CSP for configuration and analyses purposes.
We show how the detection of contradictions can be realized efficiently by
reducing the size of the \AFM, by reducing the number of checks and by making
the checks themselves more efficient.
In addition, we present how contradictions within the \AFM can be
detected automatically. To this end, we adapt the QuickXplain algorithm
\cite{Junker:2002:citeulike:2729290, Junker:2004:QPE:1597148.1597177}
in order to efficiently find a contradicting set
of constraints in the \CSP.
We implement our approach in SWI-Prolog~\cite{fruehwirth2012swi} with
the library CLP(FD), a constraint solver for finite domains written by
Triska~\cite{Triska12}. This solver provides powerful constraints for
integer arithmetic, determines and is correct ~\cite{Triska2014}. For
evaluation we use a case example from a medium-sized mechanical engineering PL with 497 features and 1990 attributes.


The outline of this paper is as follows:
In Section~\ref{sec:the_variability_model}, we describe the considered variability model
and its translation into a \CSP.
In Section~\ref{sec:contra}, the different kinds of contradictions are defined
and an efficient approach for their detection is proposed.
Afterwards, we describe in Section~\ref{sec:contra_explain} how the
contradicting relations can be found and presented to the modeler in a way that
helps him to resolve the contradiction.
A case study for the developed tool support is presented in Section~\ref{sec:case_study}.
In Section~\ref{sec:related_work}, related approaches are described, before the
paper is summarized in Section \ref{sec:conclusion}.  

\vspace{-0.5cm}
\section{The variability model}
\label{sec:the_variability_model}
\vspace{-0.3cm}

The variability of a \PL can be described by a \FM. 
To make optimization possible, each feature may have some
integer attributes.
The variability model we consider in this paper is called \AFM~\cite{DBLP:conf/caise/BenavidesTC05} if it contains features with integer-valued attributes and consists of the following
two parts:
\begin{compactitem}
  \item A (graphical) notation which represents the hierarchically arranged set
  of features and attributes with the relationships between parent- and child
  features called feature-tree.
  \item \CTCs that specify conditions, i.e., relations over the integer values of the attributes, that a product configuration must
  satisfy. These are usually written in a textual notation in Figure~\ref{fig:fmrobot}.
\end{compactitem}
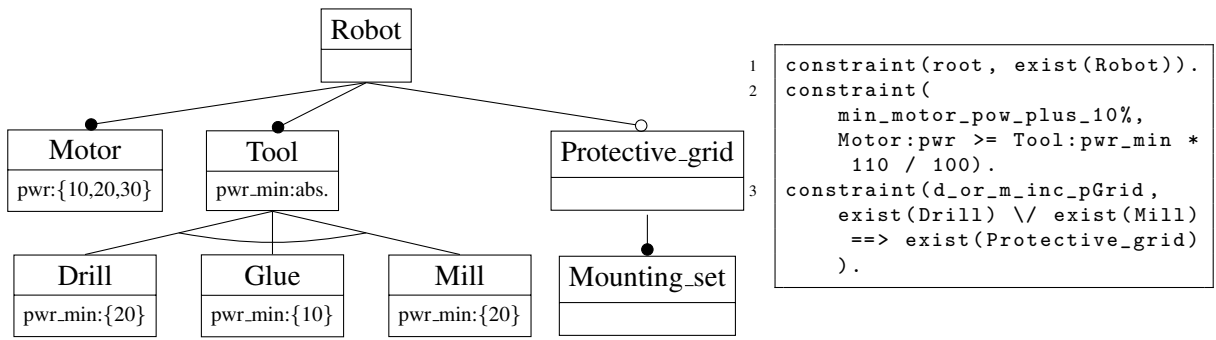
\begin{figure}[b]
\begin{minipage}{0.64\textwidth}

\begin{tikzpicture} [edge from parent path={(\tikzparentnode.south) -- (\tikzchildnode.north)},
					level 1/.style={sibling distance=7em, level distance=1.8cm},
					bend angle=10,
					scale=0.925] 
\node[rectangle split, rectangle split parts=2, draw] (robot) {Robot}
	child {
		node [rectangle split, rectangle split parts=2, outer ysep=0.6mm, draw]
		(motor) {Motor \nodepart{two}\scriptsize{pwr:\{10,20,30\}}} edge from parent
		[draw,-*] }
	child {
 		node [rectangle split, rectangle split parts=2, outer ysep=0.1mm, draw]
 		(tool) {Tool \nodepart{two} \scriptsize{pwr\_min:abs.}} edge from parent
 		[draw,-*] child {
			node [rectangle split, rectangle split parts=2, draw] (drill) {Drill
			\nodepart{two} \scriptsize{pwr\_min:\{20\}}} edge from parent [draw,-] node
			[midway] (leftmid) {} }
		child {
			node [rectangle split, rectangle split parts=2, draw] (glue) {Glue
			\nodepart{two} \scriptsize{pwr\_min:\{10\}}} edge from parent [draw,-]
		}
		child {
			node [rectangle split, rectangle split parts=2, draw] (mill) {Mill
			\nodepart{two} \scriptsize{pwr\_min:\{20\}}} edge from parent [draw,-] node
			[midway] (rightmid) {} }
  	}
	child[missing] {}
 	child {
 		node [rectangle split, rectangle split parts=2, outer ysep=0.6mm, draw]
 		(grid) {Protective\_grid} edge from parent [draw,-o]
		child {
			node [rectangle split, rectangle split parts=2, draw  ] (mounting_set)
			{Mounting\_set} edge from parent [draw,-*] }
 	};
\draw [bend right] (leftmid.center) to (rightmid.center); 
\end{tikzpicture}
\end{minipage}
\begin{minipage}{0.35\textwidth}
\begin{lstlisting}
constraint(root, exist(Robot)). 
constraint(min_motor_pow_plus_10%, Motor:pwr >= Tool:pwr_min * 110 / 100).	
constraint(d_or_m_inc_pGrid, exist(Drill) \/ exist(Mill) ==> exist(Protective_grid)).
           
\end{lstlisting}
\end{minipage}
\caption{AFM of a Robot}
\label{fig:fmrobot}
\end{figure}
As an example, Figure~\ref{fig:fmrobot} shows a feature-tree with \CTCs of a
simplified robot \PL from a real world \AFM for configuration purposes.
A \emph{Robot} consists of a \emph{Motor}, a \emph{Tool} and an optional
\emph{Protective\_grid}.
The \emph{Motor} attribute \emph{Motor:pwr} allows a motor with either $10$,
$20$ or $30$ kW driving power.
The alternative-group \emph{Tool} can be specialized either as a \emph{Drill} or
a \emph{Glue} or a \emph{Mill} tool.
Each possible tool defines its minimum input power by an attribute
\emph{pwr\_min}.
The abstract attribute \emph{Tool:pwr\_min} groups the child attributes.
The optional \emph{Protective\_grid} requires also a \emph{Mounting\_set}.
These mandatory feature chains, which do not change the variability of the
\AFM, are typical for real-world configuration models.
They reflect the hierarchically whole-part relations of a machine.
Often, they are extended by attributes like price or weight which are summarized up to
the root by \CTCs for possible optimisation.
The \CTCs are specified in a textual form by the keyword
\texttt{constraint} followed by a name and the actual constraint.
The syntax and semantics of the constraints in this case follow 
SWI-Prolog~\cite{fruehwirth2012swi}.
Feature- and attribute names are translated into variables of the constraints. 
The construct \texttt{exist} requires that the feature variable is set to true, while  the construct
\texttt{nonexist} demands the opposite.

A \FM can be translated, e.g.,  into propositional logic (SAT solver), description
logic, constraint logic, binary decision trees~\cite{Benavides:2010:AAF:1786808.1787031}.
The translation of an \AFM into a \CSP is the most suitable representation for product
configuration purposes and analyses of contradictions, because most CSP
solvers offer good support for optimisation and can handle arithmetic constraints. 
Also explaining a contradiction in terms of constraints is best done in the
same language the modeller defines the \AFM.
\begin{listing}[!t]
\begin{listingsize}
\begin{lstlisting}[frame=none]{}
Robot = Tool						// Mandatory
Robot = Motor
Protective_grid = Mounting_set
Robot = 0 ==> Protective_grid = 0	// Optional
Protective_grid = 1 ==> Robot = 1
sum([Mill, Glue, Drill], =, Tool)	// Alternative
Drill = 0 <==> Drill:pwr_min = nil	// Attribute	
Glue = 0  <==> Glue:pwr_min = nil
Mill = 0  <==> Mill:pwr_min = nil
Motor = 0 <==> Motor:pwr = nil
Tool = 0  <==> Tool:pwr_min = nil
// Abstract attribute
element(_, [Drill:pwr_min, Glue:pwr_min, Mill:pwr_min], Tool:pwr_min)
Tool:pwr_min \= 20 ==> Mill:pwr_min  \= 20
Tool:pwr_min \= 10 ==> Glue:pwr_min  \= 10
Tool:pwr_min \= 20 ==> Drill:pwr_min \= 20
// cross-tree-constraints
Robot = 1
Motor:pwr >= (Tool:pwr_min) * 110 / 100
Drill = 1 \/ mill = 1 ==> Protective_grid = 1
\end{lstlisting}
\end{listingsize}
\caption{AFM from Figure~\ref{fig:fmrobot} as constraints}
\label{lst:fmrobot_constraints}
\end{listing}
A \CSP is described by Tsang~\cite{Tsang-DBLP:books/daglib/0076790} as a Triple
$csp=\{Z, D, \mathcal{C}\}$ where $Z$ is a finite set of variables, $D$ is a set of
possible domain values for each variable and $\mathcal{C}$ is a finite, possibly empty,
set of constraints which are restricting the domain values of the variables.
A CSP solver takes a \CSP and searches for a possible solution which 
contains assignments for each variable from its corresponding
domain which satisfies all constraints.
The solving process is divided into two substantial steps.
The first step tries to reduce the solution space of the \CSP by value
propagation where some, but generally not all, invalid domain values are removed
from $D$.
In the second step, all remaining variable assignments are explored, mostly by
backtracking, to find a solution.
This step is known as the labeling process of a \CSP-Solver.
A \CSP is called consistent if a solution exists, otherwise it is called inconsistent.
We translate the \AFM into a \CSP.
The mapping of the feature-tree is shown in Table \ref{tablelabel} and is based
on the work of John~\cite{John02}.
For each feature, a variable with the possible domain values $0$ and
$1$ is created where $1$ represents the existence of the feature and $0$
its absence.
For each attribute of a feature, a variable is generated with the given possible
domain values of the attribute and an additional $nil$ value. 
The value $nil$ indicates that no domain value is valid for this attribute. 
In this case, the feature of the attribute must also be absent. 
This is captured by the constraint $P = 0 \Leftrightarrow a = nil$
which sets the value $nil$ to the attribute variable when its feature does not
exist and vice versa. Since every value of the attribute can be chosen they are
known as an optional attribute values.
\begin{minipage}{0.59\textwidth}
\begin{tabular}{|c|c|p{4.2cm}|}
\hline
\multicolumn{2}{|c|}{Relation} & CSP-Translation\\
\hline
\rotatebox{90}{\hspace{0.3cm}Feature} & 
\begin{tikzpicture}[every text node part/.style={align=center}]
\node [rectangle split, rectangle split parts=2, draw, text width=2cm] (x) at
(0,0) {$P$\nodepart{two} a:\{[1...10]\}}; \node (empty) at (0,-0.75) {};
\end{tikzpicture}
 & 
 \begin{tikzpicture}
\node [text width=4cm] { 
\begin{lstlisting}[label={lst:MandatoryCode},boxpos=t,style=nonumbers]
P in 0..1
a in 1..10 \/ nil 
P = 0  <==> a = nil
\end{lstlisting}
};
 \end{tikzpicture}\\
\hline
\begin{sideways}\hspace{0.27cm}Mandatory\end{sideways} &
\begin{tikzpicture}[every text node part/.style={align=center}, node distance=1.3cm]
\node (x) at (0,0.5) {};
\node [rectangle split, rectangle split parts=2, draw] (p) {$P$};
\node [rectangle split, rectangle split parts=2, draw, below of=p] (c) {$C$};
\draw[-*] (p) to (c);
\end{tikzpicture}
& 
\begin{tikzpicture}
\node [text width=4cm] {
\begin{lstlisting}[label={lst:OptionalCode},style=nonumbers]
C <==> P.
\end{lstlisting}
};
\end{tikzpicture}
\\
\hline
\begin{sideways}\hspace{0.43cm}Optional\end{sideways} & 
\begin{tikzpicture}[every text node part/.style={align=center}, node distance=1.3cm]
\node (x) at (0,0.5) {};
\node [rectangle split, rectangle split parts=2, draw] (p) {$P$};
\node [rectangle split, rectangle split parts=2, draw, below of=p] (c) {$C$};
\draw[-o] (p) to (c);
\end{tikzpicture} &
\begin{tikzpicture}
\node [text width=4cm] {
\begin{lstlisting}[label={lst:OptionalCode},style=nonumbers]
C = 1 ==> P = 1.
P = 0 ==> C = 0.
\end{lstlisting}
};
\end{tikzpicture}
\\
\hline
\begin{sideways}\hspace{0.3cm}Alternative\end{sideways} & 
\begin{tikzpicture}[every text node part/.style={align=center}, node distance=2cm,bend angle=15]
\node (x) at (0,0.5) {};
\node [rectangle split, rectangle split parts=2, draw] (p) at (0,0) {$P$};
\node [rectangle split, rectangle split parts=2, draw] (c1) at (-1,-1.3) {$C_1$};
\node [rectangle split, rectangle split parts=2, draw] (c2) at (0,-1.3)  {$C_2$};
\node [rectangle split, rectangle split parts=2, draw] (cn) at (1.5,-1.3)  {C$_n$};
\draw (p.south) -- (c1.north) node [midway] (leftmid) {};
\draw (p.south) -- (cn.north) node [midway] (rightmid) {};
\draw (p.south) -- (c2.north);
\draw [bend right] (leftmid.center) to (rightmid.center);
\node (dots) at (0.75,-1.3) {\ldots};
\end{tikzpicture}  & 
\begin{tikzpicture}
\node [text width=4cm] {
\begin{lstlisting}[label={lst:AlternativeCode},style=nonumbers]
sum([C1,C2..Cn],=, P).
\end{lstlisting}};
\end{tikzpicture}
\\
\hline
\begin{sideways}\hspace{0.3cm}Abstract Attribute\end{sideways} & 
\begin{tikzpicture}[every text node part/.style={align=center}, node distance=3cm,bend angle=15]
\node [rectangle split, rectangle split parts=2, draw] (p) at (0,0) {$P$ \nodepart{two} a:abstract}; 
\node [rectangle split, rectangle split parts=2,draw] (c1) at (-1,-1.5) {$C_1$ \nodepart{two} a:\{1,2\}}; 
\node [rectangle split,rectangle split parts=2, draw] (c2) at (1,-1.5)  {$C_2$ \nodepart{two} a:\{2,3\}}; 
\draw (p.south) -- (c1.north) node [midway] (leftmid) {}; 
\draw (p.south) -- (c2.north) node [midway] (rightmid) {}; \draw [bend right] (leftmid.center) to (rightmid.center); 
\end{tikzpicture}  & 
\begin{tikzpicture}
\node [text width=4cm] {
\begin{lstlisting}[label={lst:abstractAttributeCode},style=nonumbers]
element(_, [C1:a, C2:a], P:a).
P:a \= 1 ==> C1:a \= 1.
P:a \= 2 ==> C1:a \= 2 \/ C2:a \= 2.
P:a \= 3 ==> C2:a \= 3.
\end{lstlisting}};
\end{tikzpicture}
\\
\hline
\end{tabular}
\label{tablelabel}
\captionof{table}{Mapping graphical \AFM elements to a CSP}
\end{minipage}
\begin{minipage}{0.4\textwidth}
For a mandatory feature relation where the feature $C$ is a part of the feature
$P$, a constraint $P = C$ is defined.
This constraint ensures that both features or neither of them must exist.
For an optional feature $C$ which may be part of the feature $P$, the following
constraint is created $(C=1 \Rightarrow P=1) \wedge (P=0 \Rightarrow C=0)$.
This constraint ensures that, if the feature $C$ exists, the feature $P$ must
also exist and if $P$ does not exist the child feature $C$ cannot exist.
For an alternative-group where only one child feature of $P$ can exist, the
constraint \texttt{sum}$([C_{1},C_{2},...C_{n}],=,P)$ is added.
This constraint ensures that the sum of the existence variables $C_{1}...C_{n}$
is equal to the existence variables of $P$.
So neither $P$ nor any children of $P$ exist or $P$ and only one child of $P$
can exist.

For the parent feature $P$ of an alternative-group, an abstract attribute $a$
can be defined if all children $C_{1}...C_{n}$ define an attribute $a$ with the
same name and type, similar to abstract super-classes in OOP. 
The set of possible domain values for the abstract attribute $a$ at the
feature $P$ is the union of all domain values of all attributes $a$ from the
children $C_{1}...C_{n}$.
For the abstract attribute $a$ in Table 1, the domain values would be
$\{1,2,3\}$.
\end{minipage}

This can be expressed by the constraint \texttt{element}$(\_, [C_{1}{:}a,C_{2}{:}a],P{:}a)$.
If a domain value is removed from the abstract attribute, this value must also be removed from all child attributes.
A constraint of the form $P{:}a \neq dv \Rightarrow C_{i}{:}a \neq dv$ is
created for each domain value $dv$ that occurs in the attribute $a$ of $C_{i}$ 
based on John~\cite{John02}. 
John shows that this translation rules lead to a tree-like \CSP in which the nodes
represent variables and the undirected arcs represent the existence of
an constraint between them.
Freuder~\cite{Freuder:1982:SCB:322290.322292} shows that tree-like constraint graphs can be solved without any
backtracking.
So the \CSP obtained by translating the feature tree can be solved by arc-consistency~\cite{Mackworth:1975:CNR:902065} propagation
mechanism of a solver.

The \CSP generated from the \AFM in Figure \ref{fig:fmrobot} is shown in
Listing \ref{lst:fmrobot_constraints}.
Each feature name represents a \CSP variable.
In lines 1 to 3, the constraints express the mandatory features.
If \emph{Robot} exists, \emph{Motor} and \emph{Tool} must also exist, and if
\emph{Protective\_grid} exists, the \emph{Mounting\_set} must be also present.
The constraints in line 5 and 6 are derived from the optional feature \emph{Protective\_grid}.
The \texttt{sum} constraint in line 8 ensures that only one \emph{Tool} is
selected, if it exists.
In lines 7 to 11, the constraints ensure consistency between the existence
of a feature and the attribute values.
For the abstract attribute \emph{Tool:pwr\_min}, the constraints in lines 13 to 16 are generated.
The \CTCs are shown in lines 18 to 20.

\vspace{-0.5cm}
\section{Contradictions}
\label{sec:contra}
\vspace{-0.3cm}
The development of an \AFM is an error-prone task. 
Benavides et al.~\cite{Benavides:2010:AAF:1786808.1787031} propose the most
complete collection of analysis tasks we are aware of up to now.
Searching for contradictions in an \AFM is one of these analysis tasks.
It is also called detection of anomalies.
There are two kinds of contradictions that occur in a variability model:
A \textbf{model contradiction} (aka. model anomaly) is a contradiction
between the semantics of the feature tree and the \CTCs (e.g. an optional
feature which is present in all possible products due to cross-tree-constraints).
An \textbf{inconsistent model} (aka. void feature model) exists if no
product can be derived from the model at all (e.g.,~a cross-tree constraint
requires the absence of a full mandatory feature).
%
%
Model contradictions can be described as properties of the \AFM and its
elements~\cite{DBLP:conf/caise/TrinidadBC06}. The following properties can be determined automatically because these
contradictions always consist of at least one cyclic dependency of the relations
in the \AFM:
\begin{compactitem} 
  \item \textit{Void feature model}:
  This property of a \FM is true if the \FM is inconsistent which means that it
  represents no product at all.
  \item \textit{Dead feature}:
  This property is true if a feature cannot be selected in any product,
  therefore, it appears in no product.
  \item \textit{False optional feature}:
  The \textit{false optional} feature property becomes true if a feature is included in
  all products of the product line despite not being modeled as mandatory.
\end{compactitem}
In addition, these properties can be adapted to attributes domain values of \AFMs aswell.
\begin{compactitem}
  \item \textit{Dead attribute values}:
  This property of a domain value becomes true if the value cannot be assigned
  to its attribute in any product. (E.g., in Figure \ref{fig:fmrobot}, the domain
  value $10$ cannot be assigned to \emph{Motor:pwr} because a
  cross-tree constraint requires a minimum motor power of \emph{Tool:pwr\_min} plus $10$ percent). 
  This property is the same as \emph{non-attainable domain values} introduced by
  Salinesi and Mazo~\cite{salinesi:hal-00707461}.
  \item \textit{False optional attribute values}:
  An attribute value has the property \textit{false optional} if it is included in all
  products of the product line despite being modeled as an optional attribute.
\end{compactitem}
The reasons that makes any of the above properties true are always related to
cross-tree constraints.
For an \AFM without cross-tree constraints, none of these properties can be
true because the \CSP, constructed from the feature tree is a tree-like
\CSP which contains no cyclic dependencies.
Imagine a tree where the nodes are variables or values of the \CSP and the arcs
are the relation between them. 
In this case, a solution of the \CSP always exists. 
A contradiction is only possible if the tree becomes a graph by an additional
relation from a \CTC.
\vspace{-0.5cm}
\paragraph{Detection of Contradictions}
\label{sec:contra_detection}
The main operation to detect a contradiction is a method which takes an \AFM and
derives a product from it.
In other terms, a constraint solver needs to calculate a solution of a \CSP
which represents the \AFM.
This is exactly the case for detecting a \textit{void feature model}.
In order to detect the other contradictions, we need the possibility to add
additional constraints to the \AFM, e.g., to detect a \textit{dead feature}, a
constraint which selects this feature must be added.
For this reason, Trinidad \cite{Trinidad12-PhD} proposes to distinguish between
a \FM and a \SFM.
A \SFM can also contain additional constraints, corresponding to user
decisions and feature selections.
A naive approach to find contradictions is to check all features and attribute
values.
After checking the \AFM for a solution (no \textit{void feature model}), for
each feature the following steps are executed to detect contradictions: 
\begin{compactenum}
  \item Select a feature.
  A constraint is created where the variable which represent the
  feature in the \CSP is set to $1$.
  This constraint is added to a copy of the \FM resulting in a \SFM.
  \item Find a solution.
  The constraint solver is used to derive a product solution of the \SFM.
  If no solution is found, the feature is marked as a \textit{dead feature}.
  \item If the feature is optional or part of a child in an alternative-group:
  Deselect this feature.
  A constraint is created where the \CSP variable of the feature is
  set to $0$ and the constraint is added to a fresh copy of the \FM to get a
  \SFM.
  \item Find a solution.
  If no solution is found for the \SFM from the step above, this feature is
  marked as a \textit{false optional feature}.
\end{compactenum}
The process to check the attribute values is similar.
First all attributes are collected. 
For each attribute, all possible domain values must be checked. 
So for each domain value, the following steps are executed:
\begin{compactitem}
  \item Check for a \textit{dead attribute value}.
  A constraint is created which sets the attribute to the domain value, adds
  the constraint to a copy of the \AFM and searches for a solution.
  If there is no solution, this attribute value is marked as a \textit{dead
  attribute value}.
  \item Check for a \textit{false optional attribute value}.
  A constraint is created where the domain value is removed from the attribute
  and added to a copy of an \AFM.
  If no solution is found, this attribute value is a \textit{false optional
  attribute value}.
\end{compactitem}
This approach is inefficient because not all features and attribute values have
to be checked.
Some properties for certain features or attributes cannot be true because these features or attributes are not involved in cross-tree constraints.
Other properties must not be checked due to the semantics of the feature tree. E.g., a
full mandatory feature must not be checked for the \textit{dead feature}
property if the property \textit{void feature model} is false.
Additionally, some contradictory properties for certain features or attributes can be true as a side-effect if
this feature or attribute is connected to another feature or attribute by a feature-tree or cross-tree constraint
for which a contradictory property is true. 
\vspace{-0.5cm}
\paragraph{Efficient Contradiction Detection}
\label{sec:contra_detection_efficient}
To decrease the computation efforts during the search for contradictions, we
reduce the \CSP which represents the \AFM and propose additional rules to
decrease the amount of checked features.

The basic idea is to reduce the variability of the \AFM by deleting all features 
and attributes which cannot be part of any contradiction.
As a contradiction needs at least a cyclic path of its relations, we can safely
delete all leaf-features and attributes which are not part of an alternative-group
from the \AFM which do not appear in any cross-tree constraint.
Only cross-tree constraints can create cyclic relation paths in the \AFM.
In detail, we remove:
(1) Each set of attributes marked as abstract in an alternative-group where
  none of the attributes appears in any cross-tree constraint.
(2) Any other attribute which does not appear in any cross-tree constraint.
(3) The features of an alternative-group if none of the features has an
  attribute or a child-feature and none of them appears in any
  cross-tree-constraint.
(4) And any other leaf feature not appearing in any cross-tree constraint.

The analysis process benefits from the reduced \AFM in two ways:
(1) Less features and attributes have to be checked for a possible
  contradiction.
(2) A smaller \CSP can be used in the checks to search for a possible
  product.
In addition, the modeler benefits from the reduced \AFM because only the root of
a contradiction is found.
Possible aftereffects are not detected, only the head features of mandatory
feature chains are interesting for contradiction checking.
Consider the feature \emph{Mounting\_set} in Figure \ref{fig:fmrobot}.
If the feature \emph{Protective\_grid} is a \textit{dead feature}, it is
clear that \emph{Mounting\_set} is also a \textit{dead feature}. 
Therefore, it is only necessary to find and check the heads of those feature
chains.

Apart from the reduced \AFM, there are two more ways to
increase the performance of contradiction analyses.
The first is to investigate how the number of checks for a solution can be
reduced since finding a solution in a \CSP is an expensive task.
The second is to improve the solution finding process itself.
\vspace{-0.5cm}
\paragraph{Reduce Number of Checks}
While Salinesi and Mazo~\cite{salinesi:hal-00707461} only mention the possibility to reduce the amount of
solution checks by comparing the set of features to be checked against the
features of a derived product from a check, we implement this approach and improve it further as follows: 

From the reduced \AFM, we collect all optional features and children of an
alternative-group from the feature-tree into a set of features to be checked. 
This set is duplicated to a set called \texttt{check\_dead} for the \textit{dead
feature} checks and a set called \texttt{check\_false\_opt} for the
\textit{false optional feature} checks.
All features of these sets are compared with the features from the derived
product after the \textit{void feature model} check of the \AFM.
A feature from the set \texttt{check\_dead} will be removed if it is selected in
the product. Features from the set \texttt{check\_false\_opt} will be removed if
they are not in the product. 
This halves the amount of checks because each feature from the \AFM is either
selected or deselected in the derived product.
For each remaining feature of the \texttt{check\_dead} set, a \textit{dead
feature} check is performed starting with the features from the
alternative-groups where only one feature can be selected at a time.
Afterwards, the features from both sets are compared against the derived product
from the \textit{dead feature} check. 
If the \FM contains large alternative-groups, a huge amount of \textit{false
optional feature} checks can be pruned because most features of an 
alternative-group are deselected in a product.

For further reduction of \textit{dead feature} checks, we are interested in a
derived product that contains as many features as possible from the features in
the set \texttt{check\_dead}.
It can be obtained by an explicit labeling strategy for the \CSP when solving
the \textit{dead feature} check.
The variables representing the existence of the optional features from the
\texttt{check\_dead} set are labeled with a strategy which tries the highest
domain values first.
This results in a derived product where the maximum of optional features can be
removed from the set \texttt{check\_dead}.
During the \textit{false optional feature} checks, the labeling strategy tries
the smallest domain values first for the \CSP variables representing the
optional features.
During the \textit{dead feature} checks, the labeling strategy for the optional
features can be switched to the \textit{false optional feature} check strategy
if all optional features are removed from the set \texttt{check\_dead}.
Then, the \textit{dead feature} checks derive products which contain the minimum
number of optional features which are removed from the
\texttt{check\_false\_opt} set.
Depending on the amount of features in alternative-groups mostly the set of
\texttt{check\_false\_opt} features is empty when all \textit{dead feature}
checks are done.
\vspace{-0.5cm}
\paragraph{Reduce Checking Efforts}
The second way to increase performance is to speed up the CSP solver in finding a
solution.
In general, finding a solution of a \CSP is a NP-complete problem
\cite{Tsang-DBLP:books/daglib/0076790}.
With the reduced \AFM, we decrease space, not the time complexity which comes with
the cross-tree-constraints.
This leads to less propagation steps in the \CSP-Solver in the labeling process.
In practical examples the performance gain from using the reduced \AFM is about
linear to the reduced size, as the amount of propagation steps are nearly linear to the
amount of constraints generated from the feature-tree.
The labeling options of the \CSP-solver are the key to speed up solution finding.
Most solvers offer the possibility to define the order of the variables, the
domain value order and the branching strategy for the labeling process.
Unfortunately, the fastest labeling strategy depends on the specific problem.
In general, it is a good idea to store the labeling options within the \AFM.
As a starting point, we use an order where the variable with the
smallest domain is labeled first. 
Then, the detection of infeasible values happens at an early stage.
Additionally, for variables representing the existence of the features, an order
is chosen where the variable participating in most constraints are labeled
first with an enumerating branching strategy.
For attribute variables, a bisection method as a branching strategy is chosen.
These options are problem-specific and can change the
performance heavily.

\vspace{-0.5cm}
\section{Explaining Contradictions}
\label{sec:contra_explain}
\vspace{-0.3cm}

After detecting a \textit{void feature model} or an other property
indicating a contradiction, the modeler needs a clue to change this mistake in the variability model.
The essential task for removal of contradictions within the \AFM is changing
relationships between features or attribute values.
Therefore, the contradicting relations must be detected and shown to the
modeler.
Afterwards the modeler must decide which relation needs to be changed in order
to resolve the contradiction.

\begin{minipage}{0.6\textwidth}
\begin{lstlisting}{}
protective_grid = 0          % false_optional_feature
robot = 1
robot = tool                 % mandatory
sum([mill, drill], =, tool)  % alternative-group
drill = 1 #\/ mill = 1 #==> protective_grid = 1
          %drill_or_mill_include_protective_grid1 
\end{lstlisting}
\vspace{-0.5cm}
\captionof{listing}{Defiant constraints for the \textit{false optional feature}\\ \emph{Protective\_grid} if the feature \emph{Glue} is removed}
\label{lst:contradicting_constraints_fo_1}
\end{minipage}
\begin{minipage}{0.02\textwidth}
~
\end{minipage}
\begin{minipage}{0.33\textwidth}
Consider for example the \AFM shown in Figure \ref{fig:fmrobot}, if the modeler
removes the feature \emph{Glue} from the alternative-group \emph{Tool}, the
feature \emph{Protective\_grid} becomes a \textit{false optional feature}.
Listing \ref{lst:contradicting_constraints_fo_1} shows the contradicting constraints for 
\vspace{0.15cm}
\end{minipage}
the \textit{false optional feature} \emph{Protective\_grid} in this \AFM.
Line 1 shows the constraint to check the \textit{false optional feature}
property for the feature \emph{Protective\_grid}.
The feature \emph{Protective\_grid} is disabled and the \CSP becomes
inconsistent which means that the \CSP has no solution (the \AFM is a
\textit{void feature model}).
The constraint in line 2 expresses that a robot exists and the constraint in
line 3 requires the existence of the alternative-group \emph{Tool} if a robot
exists.
In line 4, the constraint determined that one feature from the alternative-group
\emph{Tool} must exist.
The \CTC in line 5 requires the feature \emph{Protective\_grid}
if the feature \emph{Mill} or \emph{Drill} exists.
Hence, the selection of feature \emph{Protective\_grid} contradicts the constraint in line 1.

For a given inconsistent \CSP, a simple approach to find the contradicting
constraints is to check all possible subsets of constraint combinations for 
consistency and select the smallest inconsistent set.
The effort for this, however, is exponential.
In real world scenarios, the modeler does not need the smallest conflict set.
An arbitrary conflict set, which has no smaller subset of conflicting
constraints, contains in most cases enough information for the modeler to solve
the conflict.
\vspace{-0.5cm}
\paragraph{QuickXPlain}
Junker developed the QuickXPlain algorithm \cite{Junker:2002:citeulike:2729290,
Junker:2004:QPE:1597148.1597177} which calculates an inconsistent set of
constraints for a given inconsistent \CSP.
The number of consistency checks for the QuickXPlain algorithm is 
$O(n,k)=2k*log_{2}(n/k) + 2k$ 
\cite{Junker:2004:QPE:1597148.1597177}, with $n$ the number of all
constraints and $k$ the number of constraints in the resulting conflict set. 
The number of constraints in the conflict set $k$ in relation to the number of
all constraints $n$ has a vital impact on the number of consistency checks.
The conflict set computed by QuickXPlain is one of possibly many, 
whereas the simple approach described above finds the smallest conflict set, but
with $O(n)=2^n$ consistency checks.
\vspace{-0.5cm}
\paragraph{Modularization of \AFM}
\label{sec:contra_modularization_of_EFM}
Finding an explanation for a contradiction is a time consuming task.
We  decrease the computation efforts by reducing the number of constraints to
check and constraints in the conflict set.
The basic idea is to call the QuickXPlain algorithm twice in a cascading way.
Therefore, we group the constraints representing the \AFM into large conjunctions
of constraints and call the QuickXPlain algorithm on this constraint-groups.
Afterwards, we split up the constraint-groups from the resulting conflict set
back into their native constraints and run QuickXPlain again on this set of
constraints.
All constraint-groups which are not part of the initial conflict found by
the first call of QuickXPlain can be ignored. This decrease the number of
constraints for the second computation.
The main challenge is using a good heuristic to build the constraint-groups for
the first call of QuickXPlain. A good gouping has a small amount of constraint-groups
and leads to small conflict set which contains in the best case with one
constraint-group.

During development the modeler maps the 
structure of the
\PL into the \AFM similar to multi software product lines~\cite{Damiani2014SPLC}. 
Thereby, the modeler encapsulates subunits, respectively assembly units, of the
\PL into \AFM-modules. That reflect the structure of the \PL into the \AFM
for separation of concerns and reuse. This existing module structure can be exploited 
as a good heuristic for grouping of the constraints for efficient explanation of
contradictions.

All constraints of a module can be consolidated to one conjunctive constraint.
It reduces the number of constraints for the first call of QuickXPlain to the
number of modules. Furthermore, the resulting conflict set is also small because most conflicts
occur within a single module or between two modules.
This grouping with cascading calls of the QuickXPlain algorithm leads to
ca.~$10$ times faster computation for the explaining conflict set (see Section
\ref{sec:case_study}).

If a module structure is not available, another idea to group the constraints is finding all clusters of connected
\CTCs and add all generated constraints from the feature-tree bottom up to these clusters.
Other groupings, e.g., by the type of the relation in the feature-tree give a
performance gain of less than 10 percent. The evaluation of different grouping approaches is left future work.

\vspace{-0.5cm}
\paragraph{Implementation Details of a modularized \AFM}
\label{sec:Implementation_Details_of_a_modularized_AFM}

In order to build an AFM-module, all external dependencies of the module can
be added as attributes in the module's root feature.
This can be done, e.g., by adding \CTCs between the root attribute and
the features or attributes internal to the module.
In our approach, we introduce feature and attribute references at the root feature which behave
like the original feature or attribute, but are accessible from other modules via
the module's root feature.
The module's root feature becomes an interface of the module, containing all
external dependencies between the features and attributes of the used module and of a super-ordinate module.
For reusable modules, we support the marking of features and
attribute values as \textit{false optional} or \textit{dead} by a
'\texttt{pragma}' directive.
This is necessary because when using one module multiple times in a greater
\AFM, the modeler does some pre-configuration of the generic module with
cross-tree constraints which leads to model contradictions. 
These intentional contradictions are ignored at
the stage of contradiction detection.

\vspace{-0.5cm}
\section{Case Study and Tool Support}
\label{sec:case_study}
\vspace{-0.3cm}


The approach developed in this paper is implemented in SWI-Prolog.
The translation of the feature-tree into constraints is written in constraint handling rules (CHR)~\cite{schr_demoen_kulchr_chr04} and SWI-Prolog.
\begin{table*}
 \centering
 \begin{tabular}{|c|c|c|c|c|c|c|c|c|c|}\hline
   Scen-& Con. & \multicolumn{4}{|c|}{Simple Search} & \multicolumn{4}{|c|}{Cascading Search} \\ \cline{3-10}
   ario & in the &\multicolumn{2}{|c|}{\AFM} & \multicolumn{2}{|c|}{reduced \AFM} & \multicolumn{2}{|c|}{\AFM} & \multicolumn{2}{|c|}{reduced \AFM} \\
    & Contra- &\multicolumn{2}{|c|}{3402 Con.} & \multicolumn{2}{|c|}{1161 Con.} & \multicolumn{2}{|c|}{$1_{st} + 2_{nd}$ search} & \multicolumn{2}{|c|}{$1_{st} + 2_{nd}$ search} \\ \cline{3-10}
    & diction & Time & Checks & Time & Checks & Time & Checks & Time & Checks \\ \hline
   s1 & 7 & 16.7 s & 61 & 5.3 s & 57 & 0.49 s & 44 & 0.47 s & 41 \\ \hline
   s2 & 2 & 7.9 s & 19 & 2.4 s & 18 & 0.42 s & 16 & 0.25 s & 11 \\ \hline
   s3 & 2 & 9.4 s & 22 & 2.1 s & 16 & 0.40 s & 15 & 0.22 s & 12 \\ \hline
   s4 & 2 & 8.4 s & 20 & 2.0 s & 15 & 0.42 s & 17 & 0.25 s & 12 \\ \hline
   s5 & 2 & 9.1 s & 21 & 2.3 s & 17 & 0.42 s & 16 & 0.31 s & 15 \\ \hline
   s6 & 7 & 8.9 s & 69 & 2.9 s & 64 & 0.87 s & 57 & 0.80 s & 61 \\ \hline
   s7 & 8 & 9.1 s & 78 & 2.8 s & 75 & 1.06 s & 67 & 0.85 s & 72 \\ \hline
 \end{tabular}
\caption{Amount of checks and execution times for finding explanations}
\label{tabletimes}
\end{table*}
As case study for evaluating our approach, we use a configuration model for generating quotes
for a medium-sized mechanical engineering application.
This model contains $497$ features with $1990$ attributes of which $237$ are optional features, $49$ alternative-groups and
$256$ cross-tree-constraints.
The feature-tree results in $3402$ constraints in the \CSP.
Using this case study, we evaluate our proposed approach for detecting
and explaining contradictions. First,
we show that the reduction of the \AFM leads to reduced computation times, both for detecting and explaining contradictions. Second, we show that the cascaded approach for explaining contradictions also decreases computation times.
Randomly generated models are not suitable for evaluation purposes as randomly generated constraints typically do not reflect realistic relationships. 

The reduced \AFM with the process described in Section~\ref{sec:contra} is about $50$ percent smaller.
It contains $222$ features with $712$ attributes, $90$ optional features, $35$
alternative-groups and $1161$ generated constraints.
In the \AFM, there are $396$ features to check for the \textit{false optional} property
($237$ optional features plus $159$ children from the alternative-groups) and
$497$ features to check for the \textit{dead feature} property.
With the efficient approach presented in Section~\ref{sec:contra_detection_efficient}, we need only $48$ consistency checks for
the remaining $161$ features to discover the contradictions.
Thereby, each consistency check is twice as fast because of the smaller amount
of constraints to propagate.
Hence, the overall performance improvement for detecting contradictions for the
reduced \AFM, compared with the native approach, is round a factor of $32$.



Table~\ref{tabletimes} shows the amount of consistency checks and the
computation time which is necessary to find a conflict set of constraints to
explain a contradiction property.
The execution times are determined on a Laptop with an Intel dual core
processor T7500 with 2200 MHz and 3 GByte RAM by the SWI-Prolog predicate
time/1 in seconds.
Each line shows different scenarios, which descibe the effort to explain a model contradiction in the \AFM.
In column 2, the number of constraints in each of the conflict sets is listed.
The block 'simple search' shows, for a single QuickXPlain call, the computation
time in seconds and the number of consistency checks.
Therein, the first two columns are for the full \AFM and the second for the
reduced \AFM.
In the second block 'cascading search' the time and checks are listed for
cascaded QuickXPlain calls for the \AFM and the reduced \AFM.
In a cascading search, the constraints are grouped by the module structure for
the first call (comp. Sect. \ref{sec:contra_modularization_of_EFM}).
These results show that our approach also improves the computation times for explaining contradictions. 
The best improvements are obtained with the cascading search. 
If we apply the cascaded search, the additional reduction of the \AFM does not matter.
If, however, the reduced \AFM is used in a single search, the time required for explaining the contradiction is improved by about
three times.

\vspace{-0.5cm}
\section{Related Work}
\label{sec:related_work}
\vspace{-0.3cm}
The analysis of \FMs is an active area of research~\cite{Batory2006,salinesi:hal-00707461}.
Most approaches consider pure feature models without attributes or just mention that their approach can be adapted to \AFMs. 
Benavides~\cite{Benavides:2010:AAF:1786808.1787031} et al. present a classification of 
different approaches for the analysis of \FMs. They mention 
numerous proposals to detect \textit{void feature models}.
Fewer approaches consider the detection of \textit{dead features}, and even less
deal with the detection of \textit{false optional features}.
In this survey only Trinidad~\cite{Trinidad12-PhD} presents an approach on how to find and explain all 
the above mentioned anomalies in a pure \FM.
For this he used a constraint solver and Reiter's theory of diagnosis. 
In contrast to our work, Trinidad does not consider integer-valued attributes and general arithmetic constraints.
There are two approaches which deal with detecting and explaining contradictions in \AFMs that are related to our work. 
For explaining contradictions in an \AFM, Zaid et al.~\cite{ZaidKT09} propose an approach based on description logic. 
This approach covers detection and explaination of \textit{dead features} using the Pellet Reasoner~\cite{peller_reasoner:Sirin:2007:PPO:1265608.1265744}.
However, in contrast to our work, they only use constraints on attribute values containing smaller-, greater- or equal-relations and no general arithmetic constraints as we do. 
%
Osman et al.~\cite{rule_based:conf/splc/OsmanPH08} detect and explain anomalies in a \FM by a knowledge-based method using propositional logic. 
He extends a \FM by so called variation points. 
The variation points are cardinalities which allow a more complex grouping of features. 
However, there is no possibility to encode attributes with the possibility to express arithmetic cross-tree constraints.

Felfernig et al.~\cite{Felfernig2012} also use constraint sets to find inconsistencies in \FMs which is very similar to our approach, but also lacks the handling of arithmetic constraints and integer-attributes.
Zhang et al.~\cite{Zhang2011ICSR} introduced refinement paths, which decreases the number of checks needed by a \SAT solver.
But they only do this on a simplied version of pure \FMs.
A more practical approach is studied by Xiong et al.~\cite{Xiong2012ICSE}, where they provide a list of solutions on encountered configuration problems in the eCos configurator.
The eCos configurator also contains numeric attributes in its configuration domain.
Their approach aims to support customers during their configuration and not the developer during modeling the \AFM.
There are other tools and languages which are designed for modeling and configuration of \AFMs.
For instance, \emph{TVL} a text-based formal language for modeling \AFMs designed by Classen et al.~\cite{Classen2011a}. 
TVL contains structuring mechanisms for \FMs, such as user-defined types or imports and \AFMs with artihmetic constraints.
Unfortunately checking such \AFMs is still under construction and currently supports only pure \FMs.
Similarly, \emph{FAMILIAR} by Acher et al.~\cite{AcherCLF13} is a textual language for representing \FMs which allows (amongst other operations) im-/ex-porting, (de-)composing, slicing and diffing of (multiple) \FMs. 
FAMILIAR supports reasoning about \FMs, but still only \FMs without attributes.
The \emph{SPLConqueror} from Siegmund et al.~\cite{SPLConqueror} uses attributed FMs and optimization functions to help stakeholders during the configuration.
They also describe the use of constraints for feature restriction, but it is unclear how errors in the model can be detected and explained to the modeler.
Another approach is the \emph{Clafer} modeling language from GSD Lab at the University of Waterloo\footnote{\url{http://www.clafer.org/}}.
It is a DSL for modeling FMs with attributes and constraints.
They support detection of void \FMs, but direct support is still missing to find e.g. dead attribute values.
Summarizing, none of the existing approaches (cf.~\cite{Benavides:2010:AAF:1786808.1787031}) supports 
the detection and explanation of contradictions, in the sense of all of the above mentioned properties, 
in \AFM with integer-value attributes and arithmetic constraints.



\vspace{-0.5cm}
\section{Conclusion}
\label{sec:conclusion}
\vspace{-0.3cm}
In this paper, we presented an approach for the detection and explanation of contradictions
in \AFMs with integer-values attributes and general arithmetic constraints. 
Based on the translation of the feature-tree into a \CSP, we presented how the
labeling process of a constraint solver can be adapted for the efficient detection of
contradictions.
The QuickXPlain algorithm~\cite{Junker:2002:citeulike:2729290,
Junker:2004:QPE:1597148.1597177} is used for computing a conflict set of
contradicting constraints which explains the reason of a contradiction.
For larger \AFMs, we used a modularization concept similar to MSPLs to improve the computation time by cascading calls of QuickXPlain.
For future work, we will further investigate possible tool integrations, heuristics to determine possible clusters of constraints in order to speed up the time for explaining contradictions.
Additionally we want to compare our approach with an implementation of the Felfering et al.~\cite{Felfernig2012} algorithm FASTDIAG combined with a diagnosis model.

%

\vspace{-0.5cm}
\nocite{*}
\bibliographystyle{eptcs}
\bibliography{lit}
\end{document}